\documentclass[11pt,reqno,notitlepage]{article}
\usepackage{setspace}
\usepackage{amssymb}
\usepackage{amsthm}
\usepackage{graphicx}
\usepackage{amsmath}
\usepackage{fancyhdr}
\usepackage{datetime}
\usepackage[toc]{appendix}
\usepackage[usenames,dvipsnames]{color}
\usepackage{tikz}
\usepackage{color}
\usepackage{colortbl}
\usepackage{mathrsfs}
\usepackage[normalem]{ulem}
\usepackage{graphicx, xcolor}
\usepackage{hyperref}
\usepackage{enumerate}

\usepackage{setspace}
\usepackage{graphicx, xcolor}												
\usepackage[mathscr]{eucal}												
\usepackage{subcaption}
\usepackage{color}
\usepackage{epstopdf}
\usepackage{ifthen}

\allowdisplaybreaks

\makeatletter
\newcommand{\labitem}[2]{%
\def\@itemlabel{\textbf{#1}}
\item
\def\@currentlabel{#1}\label{#2}}
\makeatother

\usetikzlibrary{arrows,automata,positioning}
\usepackage{pgfplots}
\pgfplotsset{compat=newest}
\usepackage[top=1in, bottom=1in, left=1in, right=1in]{geometry}
\usepackage{mathtools}									%
\mathtoolsset{showonlyrefs=true}							%

\newtheorem{theorem}{Theorem}
\newtheorem{lemma}[theorem]{Lemma}								%
\newtheorem{proposition}[theorem]{Proposition}	
\newtheorem{corollary}[theorem]{Corollary}	
\newtheorem{assumption}[theorem]{Assumption}	

\newtheorem{remark}{Remark}

\numberwithin{equation}{section}	
\numberwithin{theorem}{section}

\def\E{\mathbb{E}}		
\def\F{\mathcal{F}}		
\def\P{\mathbb{P}}										

\def\Q{\mathbb{Q}}
\def\R{\mathbb{R}}

\newcommand{\ind}[1]{\mathbb{I}_{\{#1\}}}

\newcommand{\e}[1]{\operatorname{e}^{#1}}
\newcommand{\cl}{\operatorname{cl}}
\newcommand{\co}{\operatorname{co}}
\newcommand{\bd}{\operatorname{bd}}

\DeclareMathOperator*{\argmin}{arg\,min}

\newtheorem{example}[theorem]{Example}

\long\def\symbolfootnote[#1]#2{\begingroup\def\thefootnote{\fnsymbol{footnote}}\footnote[#1]{#2}\endgroup}

\newcommand{\x}{\tilde x} 
 
\newboolean{toggleflag}
\setboolean{toggleflag}{true} 

\begin{document}

\title{The Price of Liquidity: Implied Volatility of Automated Market Maker Fees}

\author{
Maxim Bichuch
\thanks{
Department of Mathematics,
SUNY at Buffalo
Buffalo, NY 14260. 
{\tt mbichuch@buffalo.edu}. Work is partially supported by Stellar Development Foundation
Academic Research Grants program. Work is partially supported by NSF grant DMS-2420974.}
\and  Zachary Feinstein
\thanks{
School of Business,
Stevens Institute of Technology,
Hoboken, NJ 07030, USA,
{\tt  zfeinste@stevens.edu}. Work is partially supported by Stellar Development Foundation
Academic Research Grants program.}
}
\date{\today}
\maketitle

\begin{abstract}
An automated market maker (AMM) provides a method for creating a decentralized exchange on the blockchain. For this purpose, individual investors lend liquidity to the AMM pool in exchange for a stream of fees earned from its operations as a market maker. Within this work, we reinterpret the loss-versus-rebalancing as the implied fee stream generated by an AMM so that a risk-neutral investor is indifferent in the decision of providing liquidity. With this implied fee structure, we propose a novel fixed-for-floating swap on the fees generated by an AMM in order to quote the implied volatilities and implied correlations of digital assets. We apply this theory to realized fees in different markets to empirically validate the relevance of the deduced fee-based volatility.
\\
{\bf Keywords:} Decentralized finance, automated market maker, loss-versus-rebalancing, implied volatility, fixed-for-floating swap
\end{abstract}

\section{Introduction}\label{sec:intro}

Decentralized Exchanges (DEXes) provide a new approach to create financial markets using smart contracts on the blockchain. 
The primary DEX designs, which maintains true on-chain decentralization, are the so-called Automated Market Makers (AMMs). 
These contracts allow investors (liquidity providers (LPs)) to lend their securities to act as a market maker in return for fees paid by traders (liquidity takers).
However, these market designs are far from perfect; AMMs fail to satisfy all the desired market making properties as is shown in \cite{park2023conceptual}.

As in~\cite{bichuch2024defi,chen2023liquidity,hasbrouck2025economic}, we consider an LP's investment in an AMM (i.e., providing liquidity) as a derivative product which earns a dividend stream (fees) during the life of the position. In this way, the optimal strategy of the risk-neutral LP is to find the withdrawal time that maximizes his or her expected (discounted) revenue.
However, in contrast to those works which use discrete-time models, we take the continuous-time approach of, e.g., \cite{cartea2023predictable,cartea2024decentralized,milionis2022automated,capponi2025optimal,bergault2025optimal} in order to study the LP's optimal strategy. 
Similar to our approach herein, \cite{milionis2022automated} find the losses that LPs would incur relative to the self-financing portfolio that replicates the holdings of the AMM without fees. However, we consider the converse problem of the loss-versus-rebalancing by positing such losses would, in the risk-neutral measure, need to be compensated exactly by the earned fees in expectation. Functionally, \cite{li2025implied} also considers this inverse problem by relating the risk-neutral expected impermanent loss to volatility.

Because of the aforementioned idea that an investment in an AMM can be thought of as a derivative, we are specifically motivated by the question of how to quote an implied volatility from these decentralized instruments. For this purpose we first utilize the loss-versus-rebalancing as the \textbf{implied fee rate} for a generic AMM, i.e., the fee structure that is implied by the cost of the liquidity token of a specific AMM pool and the underlying (risk-neutral) price processes. Notably, we find that these implied fees are consistent between any AMM constructions.
These implied fees are in contrast to the prior literature on optimal liquidity provision which assumes \emph{ex ante} a known structure to the fees; we refer the interested reader to, e.g., \cite{cartea2024decentralized,capponi2025optimal,bergault2025optimal} in which fees are collected at a constant rate in time. 
Furthermore, our approach is distinct from the recent literature (see, e.g., \cite{he2024optimal,baggiani2025optimal}) which has focused on specific structural forms for \emph{optimal} fees.
Using these implied fees, we introduce the idea of a \textbf{fixed-for-floating swap} with the fees earned by the LP as the floating leg. Via the implied fees, the \textbf{implied volatilities and correlations} of assets can be derived. 
On empirical data, we compare these fee-based volatilities to realized historical and implied volatility measures; due to the historical nature of the fees, we find a strong equivalence between the fee volatility and historical volatility. However, we conjecture that this relation would, instead, hold for the implied volatility if data from a forward-looking fixed-for-floating swap was found.

The organization of this paper is as follows. In Section~\ref{sec:motivation}, we empirically study the fees that are collected by LPs in a constant product market maker; in doing so, we find strong statistical evidence that LPs are being compensated for their loss-versus-rebalancing. This evidence motivated us to study the theoretical relationship for risk-neutral investors. The background and notation for AMMs is provided in Section~\ref{sec:amm}. Within Section~\ref{sec:fees}, we present the optimal investment problem for the risk-neutral LP. In this way, we derive the implied fee structure for the risk-neutral investor based on the cost of investing in an AMM. These implied fees are applied in Section~\ref{sec:vol} in order to derive the resulting implied volatilities and correlations based on a novel fixed-for-floating fee swap that we introduce. In Section~\ref{sec:cs}, we apply these implied volatility formulas to empirical data to compare the implied, realized, and fee-based volatility measure. Section~\ref{sec:conclusion} concludes. All proofs are relegated to Appendix~\ref{proofs}.

\section{Motivating Example}\label{sec:motivation}

Fundamentally, an AMM is simply a pool of assets which is willing and able to act as a counterparty to any trader. Distinct from traditional market making, individual LPs capitalize the AMM's pool with physical assets against which traders transact. These transactions are executed at algorithmically determined prices which incorporate appropriate price impacts. However, central to their construction as \emph{passive} market makers, the LPs will lose value on every trade relative to the replicating portfolio in a frictionless market (i.e., the value the LP would have had if they rebalanced their portfolio on an external market to match the composition of the AMM); this cost was first defined in \cite{milionis2022automated} and was termed the \emph{loss-versus-rebalancing} (LVR). However, as a market maker, the LPs will also earn fees on every transaction which needs to be sufficient to compensate for (or exceed) the realized LVR.

As highlighted in the introduction, one goal of this work is to explicitly study the expected fees earned by LPs in an AMM. Before considering the theoretical aspects of that problem, we want to empirically examine fees earned in different pools. For simplicity, in this construction we assumed that the LP of interest invested in the constant product market maker (CPMM) which is formally defined below in Example~\ref{ex:cpmm-R}.
\begin{itemize}
\item First, we examine the WETH/USDC (5bps fee) Uniswap v3 pool on the Ethereum blockchain\footnote{Smart contract: 0x88e6a0c2ddd26feeb64f039a2c41296fcb3f5640} for the calendar years 2023-2024. We consider an LP who provided liquidity over the entire price range, effectively mimicking a CPMM. Assuming that this LP's liquidity does not alter the price path (including intra-block price changes) in the pool, the realized fees for a fixed initial investment on January 1, 2023 at midnight can be found at every block. Similarly, the realized LVR (assuming rebalancing occurs at the end of each block at the pool's spot price) can be replicated. Summing over a 30 day interval (rolling 1 block at a time) to find both the realized fees and realized LVR for an LP who invests for exactly that time interval, we find that the fees and LVR are 90.05\% correlated. In fact, with this strong linear relationship, we find that the realized LVR is approximately 2.94\% lower than the realized fees for these overlapping 30 day intervals (i.e., LVR $\approx 0.9706 \times$ fees).
\item Second, we consider a simulated CPMM with 30bps fees on SPY data from 2023.\footnote{30bps fees were chosen to match the fees collected in Uniswap v2.} Herein we collected bid-ask data at 1-second intervals and simulated transactions in the pool based solely on stale price arbitrage against the high bid and low ask during each interval.\footnote{If there is arbitrage against both the high bid and low ask, the order of transaction is determined to minimize the fees collected during that 1-second interval.} In this way both the fees and LVR (assuming rebalancing occurs at the end of each second at the closing bid or ask price depending on the sign of the rebalancing) can be simulated. As with the blockchain data, we sum over a 30 day interval (rolling 1 second at a time) to find both the realized fees and realized LVR for an LP who invests for exactly that time interval. Empirically the fees and LVR are 99.86\% correlated such that LVR is approximately 1.95\% greater than the realized fees for these overlapping 30 day intervals (i.e., LVR $\approx 1.0195 \times$ fees).
\end{itemize}

These empirical results motivate the rest of this work. Specifically, in Section~\ref{sec:fees}, we use the definition of the LVR in order to thereotically justify its close relationship to the realized fees. Then, in Section~\ref{sec:vol}, we use this relationship to construct a pricing mechanism for a novel fixed-for-floating fee swap which can be used to help LPs hedge their exposures.

\section{Automated Market Makers}\label{sec:amm}

As conceptually described above in Section~\ref{sec:motivation}, an AMM is essentially a pool of $n \geq 2$ assets against which any investor can transact. Distinct from traditional market making, AMM's determine prices by simple mathematical formulas and, in exchange, earn fees for this service. 
Following the geometric interpretation of \cite{angeris2023geometry}, an AMM can be characterized by a set of \emph{reachable portfolios} 
\[R \in \{S \subseteq \R^n_+ \; | \; S = \cl\co(S + \R^n_+),~ S \neq \emptyset\},\]
where $\cl$ defines the closure operator, $\co$ defines the convex hull, and addition is defined via the Minkowski sum. That is, an AMM can be characterized as a closed, convex, nonempty set $R$ such that if $r \in R$ and $\bar r \geq r$ (component-wise) then $\bar r \in R$.
The reachable set $R$ is the collection of portfolios that the AMM will allow itself to attain; if a trade were to push the AMM's position outside of this set, then the trade would be rejected.
The initial investors in the AMM, who provide the assets for the initial portfolio $r \in R$ are often referred to as liquidity providers or LPs.
\begin{assumption}\label{ass:R-bd}
For the remainder of this work, we will assume $R - r \not\subseteq \R^n_+$ for every $r \in \R^n_+ \backslash \{0\}$. This corresponds to the concept of No Wasted Liquidity from \cite{bichuch2022axioms}, i.e., the entire market liquidity can be purchased for a, potentially infinite, cost.
\end{assumption}
\begin{example}\label{ex:cpmm-R}
The most frequently cited AMM is the \emph{constant product market maker} (CPMM). The CPMM can be characterized by the reachable set $R_{CPMM}^L := \{r \in \R^n_+ \; | \; \prod_{i = 1}^n r_i \geq L^n\}$ for some constant $L > 0$. The level $L$ characterizes the liquidity available in the CPMM, i.e., the AMM holds larger portfolios when $L$ is larger. Throughout this work, we will sometimes refer $L$ as the number of liquidity tokens of the CPMM.
\end{example}
\begin{assumption}\label{ass:R-int}
Though it is permissible for a trader to move the AMM into the interior of its reachable set $R$, doing so means that the trader is acting suboptimally.
For the remainder of this work, we assume the AMM's pool position remains on the boundary of its reachable set $\bd R$.
\end{assumption}
\begin{assumption}\label{ass:n=2}
For the remainder of this work, we will consider AMMs on only $n = 2$ assets. We will call these assets $x$ and $y$ respectively. We will also arbitrary choose the second asset to be the num\'eraire asset within the AMM.
\end{assumption}
Following Assumptions~\ref{ass:R-int} and~\ref{ass:n=2}, any AMM defines a parametric curve $(x,y): \R_{++} \to \bd R$. Without loss of generality, we can interpret this parametrization via the (relative) price $q$ of $x$ denominated in units of $y$. Following, e.g.,~\cite{angeris2020improved,angeris2021replicating,milionis2022automated}, we define the \emph{pool value function}  by the optimization problem: $\bar C(q) = \min_{(x,y) \in R} (qx + y)$ at price $q > 0$. The pool value function provides the value (in terms of the num\'eraire asset $y$) of the AMM's pool when arbitrageurs extract as much profits from the AMM as possible. By construction of the reachable set $R$, there always exists an optimal solution $(x(q),y(q))$ to the pool value function for every $q > 0$; through this optimization, we can thus characterize the parametric curve $q \in \R_{++} \mapsto (x(q),y(q)) \in \argmin_{(x,y) \in R} (qx + y)$.
\begin{example}\label{ex:cpmm-xy}
Consider the CPMM introduced in Example~\ref{ex:cpmm-R}. This AMM can be characterized by $x_{CPMM}^L(q) := L/\sqrt{q}$ and $y_{CPMM}^L(q) := L\sqrt{q}$ with $\bar C_{CPMM}^L(q) := 2L\sqrt{q}$ for every $q > 0$ where $L > 0$ defines the liquidity available.
\end{example}
In the following lemma, we discuss how the AMM can be completely characterized through the selectors $q\mapsto(x(q),y(q))$. In fact, such an approach is used to define AMMs in, e.g., \cite{milionis2022automated}. This formulation of AMMs will be utilized throughout the rest of the paper
\begin{lemma}\label{lemma:amm}
Given an AMM, there exists \emph{portfolio update functions} $x: \R_{++} \to \R_+$ nonincreasing and $y: \R_{++} \to \R_+$ nondecreasing such that $q x(q) + y(q) = \inf_{p > 0} (q x(p) + y(p))$ for every $q > 0$.
Conversely, any such portfolio update functions $(x,y): \R_{++} \to \R^2_+$ define an AMM.
\end{lemma}
\begin{assumption}\label{ass:differential}
For the remainder of this work, we will characterize AMMs by the portfolio update functions $q \mapsto (x(q),y(q))$ satisfying the conditions of Lemma~\ref{lemma:amm}.
In addition, for the remainder of this work, we will assume that $x: \R_{++} \to \R_+$ and $y: \R_{++} \to \R_+$ are twice differentiable on $\{q > 0 \; | \; (x(q),y(q)) \in \R^2_{++}\}$.
Without loss of generality, we consider $x'(q) = y'(q) = 0$ for any $q > 0$ such that $\min\{x(q),y(q)\} = 0$.
\end{assumption}
\begin{corollary}\label{cor:amm-diff}
Under Assumption~\ref{ass:differential}, an AMM is defined by the mapping $q \in \R_{++} \mapsto (x(q),y(q)) \in \R^2_+$ such that
\begin{itemize}
\item $x'(q) \leq 0$ and $y'(q) \geq 0$ for every $q > 0$; and 
\item $q x'(q) + y'(q) = 0$ for every $q > 0$. 
\end{itemize}
\end{corollary}
\begin{remark}\label{rem:barC}
Following the characterization of AMM's from Corollary~\ref{cor:amm-diff}, it can be shown that the pool value function $q \in \R_{++} \mapsto \bar C(q) = q x(q) + y(q)$ is nondecreasing and concave. As in \cite[Lemma 5.4]{bichuch2022axioms}, this immediately implies nonnegative divergence loss for any AMM, i.e., $q x(q_0) + y(q_0) \geq q x(q) + y(q)$ for every $q,q_0 > 0$; in fact $q \mapsto [q x(q_0) + y(q_0)] - \bar C(q)$ is convex and nonincreasing for $q < q_0$ and nondecreasing for $q > q_0$.
\end{remark}

\begin{assumption}\label{ass:strict}
For the remainder of this work, we will assume that there exist prices $0 < \underline{q} < \overline q$ such that $(\underline q,\overline q) \subseteq \{q > 0 \; | \; (x(q),y(q)) \in \R^2_{++}\}$.
Furthermore, we will assume that $x'(q) < 0$ and $y'(q) > 0$ for any $q \in (\underline q,\overline q)$.
\end{assumption}

Utilizing the portfolio update functions, we can provide a characterization of price impacts experienced by traders on the AMM via the curvature of this parametrized curve. Comparable to the price impact oracle defined in \cite[Definition 3.17]{bichuch2022axioms}, we consider the curvature of the AMM:
\[\kappa(q) = \frac{x'(q)y''(q) - y'(q)x''(q)}{(x'(q)^2 + y'(q)^2)^{3/2}}, \quad q > 0.\]
In fact, by matching curvatures, we are able to (locally) map the amount of liquidity available in different AMMs. In particular, due to its prevalence, throughout this work, we will benchmark AMMs via the CPMM (see Examples~\ref{ex:cpmm-R} and~\ref{ex:cpmm-xy}). In particular, this notion allows us to match how AMMs collect fees as we vary the structure of the portfolio update functions.
The following proposition formalizes the idea in \cite[Footnote 12]{milionis2022automated} that AMM invariants can be directly compared via the curvature of their bonding function.
\begin{proposition}\label{prop:L}
Consider an AMM $(x(\cdot),y(\cdot))$. The price impacts of this AMM at price $q > 0$ are equivalent to the CPMM with
\begin{equation}\label{eq:L}
L(q) = -2 q^{3/2} x'(q) 
\end{equation}
liquidity available.
\end{proposition}

\section{Loss Versus Rebalancing and the Implied Fee Structure}\label{sec:fees}

As discussed in Section~\ref{sec:amm} above, when entering a liquidity position, the LP provides the physical units $(x(q),y(q)) \in \R^2_+$ at relative price $q > 0$ between the assets. Similarly, when leaving this position, the LP recovers the physical units $(x(q),y(q)) \in \R^2_+$. Therefore, given \emph{dollar-denominated} prices $(p_x,p_y) \in \R^2_{++}$ for these two assets respectively, the price of the liquidity position is given by \[C(p_x,p_y) := p_x x(p_x/p_y) + p_y y(p_x/p_y) = p_y \bar C(p_x/p_y).\]
However, once trading commences, this portfolio is not the total wealth for the LP. By investing in the AMM, LPs collect fees from the market making operations of the AMM. Within this section, our goal is to deduce a risk-neutral fee structure for AMMs.

\begin{assumption}\label{ass:gbm}
For the remainder of this work, we will assume 2 price processes $(P^x_t),(P^y_t)$ that follow correlated, risk-neutral, geometric Brownian motions:
\begin{align*}
dP^x_t &= P^x_t[r dt + \sigma_x dW_t], \quad P^x_0 > 0,\\
dP^y_t &= P^y_t[r dt + \sigma_y (\rho dW_t + \sqrt{1-\rho^2}d\bar{W}_t)], \quad P^y_0 > 0,
\end{align*}
for risk-free rate $r \geq 0$, volatilities $\sigma_x,\sigma_y \geq 0$, correlation $\rho \in [-1,1]$, and two dimensional Brownian motion $(W,\bar{W})$ on a filtered probability space $(\Omega,\F,(\F_t)_{t\ge0},\P)$.
\end{assumption}

\begin{remark}\label{rem:sigmaB}
We wish to highlight that $\sigma_x,\sigma_y \geq 0$ allows for the case in which one asset has zero volatility. This allows us to consider the case for, e.g., tokenized money market accounts as a stablecoin. Further discussion of stablecoins are provided in Remark~\ref{rem:stablecoin}.
\end{remark}

Recall that the goal of this section is to deduce a risk-neutral fee structure for AMMs. For that purpose, we assume that fees are collected continuously in time at a state-dependent, time-homogeneous rate. As the system is driven entirely by the prices $(P^x_t),(P^y_t)$, we will consider fees that grow according to:
\[d(fees)_t = F(P^x_t,P^y_t)dt\]
for all times $t \geq 0$ for some function $F: \R^2_{++} \to \R_+$.
In this way, we can consider $F(p_x,p_y)dt$ to be the instantaneous \emph{dollar-denominated} fees that are collected when the prices are $(p_x,p_y) \in \R^2_{++}$. Note that this construction admits any state-dependent and time-predictable fee collection model. 
As hinted at empirically in Section~\ref{sec:motivation} and will be demonstrated in Theorem~\ref{thm:fees} and Corollary~\ref{cor:indifference} below, these risk-neutral fees are directly related to the loss-versus-rebalancing (LVR) as introduced in \cite{milionis2022automated}; notably, that work presented the LVR under $r = 0$ and $\sigma_y = 0$ whereas we allow these to take positive values as well.

Every LP now has a choice of when to exit their liquidity position. This naturally leads to an optimal stopping problem. Assuming that the LP is risk-neutral, he or she attempts to find the stopping time that maximizes the aggregate fees plus the (discounted) withdrawal value of the AMM. Mathematically, the LP has a value $V(P^x_0,P^y_0)$ of
\begin{equation}\label{eq:value}
V(P^x_0,P^y_0) := \sup_{\tau \in \mathcal{T}} \E\left[\int_0^\tau e^{-rt} F(P^x_t,P^y_t)dt + e^{-r\tau}C(P^x_\tau,P^y_\tau)\right]
\end{equation}
at initial prices $(P^x_0,P^y_0) \in \R^2_{++}$ when optimizing over the set of \emph{bounded} stopping times $\mathcal{T}$. 

\begin{theorem}\label{thm:fees}
Let $(P^x_t),(P^y_t)$ follow Assumption~\ref{ass:gbm}.
Set $V_t := \int_0^t e^{-rs} F(P^x_s,P^y_s)ds + e^{-rt}C(P^x_t,P^y_t)$ for every time $t \geq 0$.
Then $V$ is a martingale if and only if $F(p_x,p_y) = p_y \ell(p_x/p_y)$ for any $(p_x,p_y) \in \R^2_{++}$ where $\ell$ is the instantaneous LVR, i.e., $\ell(q) := -\frac{1}{2}(\sigma_x^2 - 2\rho\sigma_x\sigma_y + \sigma_y^2)q^2x'(q) = \frac{1}{2}(\sigma_x^2 - 2\rho\sigma_x\sigma_y + \sigma_y^2)qy'(q)$ for every $q > 0$. 
\end{theorem}

Theorem~\ref{thm:fees} implies that the risk-neutral LP will be indifferent between all stopping times under the \emph{implied} fee structure $(p_x,p_y) \in \R^2_{++} \mapsto p_y \ell(p_x/p_y)$. 
In this way, at all times, the risk-neutral LP is indifferent between withdrawing or continuing to hold his or her liquidity position.
This is formalized in the following corollary.
\begin{corollary}\label{cor:indifference}
Let $F(p_x,p_y) = p_y \ell(p_x/p_y)$ for any $(p_x,p_y) \in \R^2_{++}$ as defined in Theorem~\ref{thm:fees}.
A risk-neutral LP is indifferent on all stopping times, i.e., $V(P^x_0,P^y_0) = \E[\int_0^\tau e^{-rt}F(P_t^x,P_t^y)dt + e^{-r\tau}C(P_\tau^x,P_\tau^y)]$ for every $\tau \in \mathcal{T}$.
Furthermore, and as a direct consequence, $V(P^x_0,P^y_0) = C(P^x_0,P^y_0)$ for any $(P^x_0,P^y_0) \in \R^2_{++}$.
\end{corollary}

\begin{remark}\label{rem:impermanent}
Corollary~\ref{cor:indifference} directly relates the LVR to the (risk-neutral) expected impermanent loss when measured over equivalent time periods. Thus, though proposed separately, they are can be seen as capturing the same risks. This relation is also seen in, e.g., \cite[Proposition 3.1]{li2025implied}.
\end{remark}

\begin{remark}\label{rem:numeraire}
The implied fee structure $(p_x,p_y) \in \R^2_{++} \mapsto p_y \ell(p_x/p_y)$ provides the interpretation that $\ell(q)$ is a fee rate provided in the second (num\'eraire) asset when the relative price in the AMM is $q > 0$. 
In fact, if the AMM is symmetric between the two assets (i.e., $x(q) = y(1/q)$ for every $q > 0$), then we recover that $\ell(q) = q \ell(1/q)$ for every price $q > 0$. In this way, under the change of num\'eraire to the first asset, the overall fee structure is unchanged since $p_y \ell(p_x/p_y) = p_x \ell(p_y/p_x)$. Therefore, under a symmetric AMM, it becomes clear that the choice of num\'eraire asset is arbitrary.
\end{remark}

From Theorem~\ref{thm:fees}, we can characterize the \emph{implied} fee structure via the instantaneous LVR $\ell: \R_{++} \to \R_+$. The following corollary considers the consistency of this fee structure across different AMM designs. Specifically, we find that this fee structure matches that of the CPMM under the variable liquidity token structure introduced in~\eqref{eq:L}.
\begin{corollary}\label{cor:cpmm}
For any AMM, $\ell(q) = L(q)\ell_{CPMM}(q)$ where $\ell_{CPMM}(q) = \frac{\sigma_x^2 - 2\rho\sigma_x\sigma_y + \sigma_y^2}{4}\sqrt{q}$ is the implied fee structure for a single liquidity token of the CPMM ($L = 1$).
\end{corollary}

\begin{remark}\label{rem:stablecoin}
All results presented within this section assume that both assets follow the risk-free drift $r \geq 0$. However, in practice, most stablecoins are constructed to match the value of fiat currency, e.g., US dollars, without accounting for the time-value of money even under strictly positive risk-free rates $r > 0$. In fact, repeating the above calculations in which $dP^y_t = 0$ with $r > 0$ introduces arbitrage between different AMMs as the implied fee structures will \emph{not} match at all prices $q$ as was found in Corollary~\ref{cor:cpmm}.

To make this concept more concrete, we can compare the instantaneous LVR for the CPMM to that of a concentrated liquidity position on $[p_L,p_U]$ on a pool of a risky asset and a stablecoin. 
\begin{itemize}
\item \emph{Constant Product Market Maker}: Following the logic in the proof of Theorem~\ref{thm:fees} where $C_{CPMM}(P_t^x,1) = 2L\sqrt{P_t^x}$ for some $L > 0$, we recover:
    \begin{align*}
    d(e^{-rt}C_{CPMM}(P_t^x,1)) &= e^{-rt}L\left[-2r\sqrt{P_t^x} + rP_t^x\left(\frac{1}{\sqrt{P_t^x}}\right) + \frac{1}{2}\sigma_x^2 (P_t^x)^2\left(-\frac{1}{2(P_t^x)^{3/2}}\right)\right]dt \\ &\qquad\qquad + e^{-rt}L\sigma_x P_t^x\left(\frac{1}{\sqrt{P_t^x}}\right)dW_t \\
        &= -e^{-rt} L \Big[\underbrace{\left(r + \frac{\sigma_x^2}{4}\right)\sqrt{P_t^x}}_{=:\;\ell_{CPMM}(P_t^x)}dt - \sigma_x\sqrt{P_t^x}dW_t\Big].
    \end{align*}
    That is, the instantaneous LVR for a single liquidity token is given by $\ell_{CPMM}(q) = (r + \frac{\sigma_x^2}{4})\sqrt{q}$ for any price $q > 0$. 
\item \emph{Concentrated Liquidity}: Following the same approach as in the CPMM case, for $L_{[p_L,p_U]}$ liquidity tokens leveraged on $[p_L,p_U]$, a calculation shows that $\ell_{[p_L,p_U]}(q) = L_{[p_L,p_U]} [(r + \frac{\sigma_x^2}{4})\sqrt{q} - r \sqrt{p_L}] \ind{q \in [p_L,p_U]}$. 
\end{itemize}
In particular, for $r > 0$ and $p_L > 0$, it  follows that $\ell_{[p_L,p_U]}(q) < L_{[p_L,p_U]} \ell_{CPMM}(q)$ for any $q \in [p_L,p_U]$. As a direct result, if the fees are earned based on $\ell_{CPMM}$, then the concentrated liquidity position will result in a submartingale and long-run profits in excess of the risk-free rate.
For the remainder of this work, we will consider any stablecoin to be a tokenized money market account to sidestep this problem. 
\end{remark}

\section{Implied Volatility and Correlation}\label{sec:vol}
Following the implied fee structure of Theorem~\ref{thm:fees}, if it was possible to calibrate the instantaneous LVR $\ell$ to data, then it would be possible to quote implied volatilities and correlations for digital assets via the construction of decentralized exchanges. Within this section, we propose a fixed-for-floating swap on the fees of an AMM which would permit this. In the following, we will first describe both the floating and fixed legs of this swap and, then, provide a discussion on how this product can be used to quote implied volatilities. While this swap is novel for AMMs, the fundamental construction is comparable to the swapping of yield tokens in Pendle Finance for lending platforms.

In constructing the fixed-for-floating fee swap, we need to carefully define the payments that each leg needs to make.
The fixed leg provides cash (or stablecoins) at the initial execution time of this swap. In enforcing the payment to be made at the execution time, there is no counterparty risk for the provider of the floating leg.
The floating leg will provide the agreed upon number of liquidity tokens of the relevant AMM (e.g., the CPMM) which will be locked up for the duration of the swap, i.e., placed in escrow. For this purpose, the fees collected -- typically a fraction of the assets being sold to the AMM (see, e.g., \cite{angeris2020improved,lipton2021automated}) -- are provided directly to the provider of the fixed leg. As the liquidity token is locked up, there is no counterparty risk for the provider of the fixed leg either. As neither counterparty is subject to default risks, this fixed-for-floating fee swap can be encoded in a smart contract to be executed on the blockchain directly. 

\begin{remark}
Though we propose herein that the fixed leg is paid in a stablecoin even for a swap with a pair of risky assets, the following discussion holds similarly with a generic asset posted as the fixed leg as well.
\end{remark}

To construct this proposed smart contract, there are two key elements that need to be specified: (i) the duration of the swap and (ii) the clearing price for the fixed leg of the swap.
First, we propose that these fixed-for-floating fee swaps provide a standardized length $T > 0$ so as to reduce market fragmentation and not require clearing over both price and time. 
As an example, on the Ethereum blockchain, $T$ can be defined as a fixed multiple of epochs (each Ethereum epoch is made up of 32 blocks or 6.4 minutes). 
With the standardized length of the contract, the clearing price for the fixed leg needs to be determined. Herein, we propose implementing a batch auction in which AMM liquidity providers will post the minimum \emph{fixed} price they would need to be paid per liquidity token in order to engage in the swap as well as the number of liquidity tokens they are offering; the providers of the fixed leg will, in turn, post the maximum price they are willing to pay for the fee stream of a liquidity token along with the desired number of liquidity tokens.  Standard batch auction clearing (see, e.g., \cite{budish2014implementation}) results in a single fixed leg which clears as many orders as possible. 

Given this fixed-for-floating fee swap, assume the clearing fixed leg for a single liquidity token on the AMM is $\bar\pi > 0$. We want to construct the implied volatilities and correlations for digital assets via the implied fee structure of Theorem~\ref{thm:fees}. 
Following risk-neutral valuation, we can find that $\bar\pi$ must be the expected value of the implied fees over the course of the contract. That is, we are searching for $\sigma_x^*,\sigma_y^*,\rho^*$ so that $\bar\pi = \E[\int_{0}^{T} e^{-rt} P_t^y \ell(P_t^x/P_t^y) dt]$ when $\bar\pi$ is being paid by the fixed leg. 

To limit the number of variables, we will first consider the case in which $y$ is a tokenized money market account so that $\sigma_y = 0$ and, without loss of generality, $\rho = 0$ as well. In this way, we are only looking for the implied volatility $\sigma_x^*$ of $x$.

\begin{theorem}\label{thm:vol}
Consider an AMM pool between a risky asset ($x$) and a tokenized money market account ($y$). 
Without loss of generality, assume $P_0^y = 1$.
There exists a bijective relation between the fixed leg of the swap $\bar\pi \in [0,\bar C(P_0^x))$ and the \emph{implied} volatility $\sigma_x^* \geq 0$ such that $\bar\pi = \E[\int_0^T \ell(e^{-rt}P_t^x)dt]$.
\end{theorem}

\begin{remark}\label{rem:arb-vol}
The upper bound $\bar\pi < \bar C(P_0^x)$ for the fixed leg of the swap follows, also, from a simple arbitrage argument. Specifically, as the liquidity token can be purchased for $\bar C(P_0^x)$, an investor could use the fixed-for-floating fee swap to obtain a risk-free profit if $\bar\pi \geq \bar C(P_0^x)$.
\end{remark}
\begin{example}\label{ex:cpmm-vol}
Consider the CPMM introduced in Example~\ref{ex:cpmm-xy} with $L = 1$ in which the num\'eraire asset ($y$) denotes the money market account with $P_0^y = 1$. The implied volatility $\sigma_x^*$ for the risky asset in the CPMM, given a price of $\bar\pi > 0$ for the fixed leg of the swap, is given by:
\[\sigma_x^{CPMM} = \sqrt{\frac{8}{T}\log\left(\frac{2\sqrt{P_0^x}}{2\sqrt{P_0^x}-\bar\pi}\right)}\]
for $\bar\pi \in [0,2\sqrt{P_0^x})$.
\end{example}

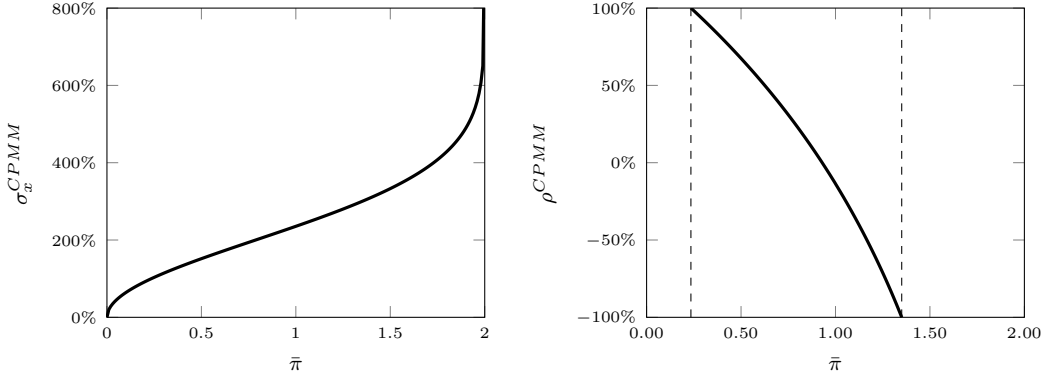
\begin{figure}[t]
\centering
\begin{subfigure}[t]{0.4\textwidth}
\centering
\begin{tikzpicture}[declare function={sigma(\x)=(8*ln(2/(2-\x)))^0.5;},]
\begin{axis}[
width={\textwidth},
xlabel={$\bar\pi$},
ylabel={$\sigma_x^{CPMM}$},
xmin=0, xmax=2,
ymin=0, ymax=8,
samples=200,
yticklabel={\pgfmathparse{\tick*100}\pgfmathprintnumber{\pgfmathresult}\%},
label style={font=\scriptsize},
ticklabel style={font=\tiny},
]
\addplot [domain=0:2,solid,color=black,very thick]{sigma(x)};
\end{axis}
\end{tikzpicture}
\caption{Example~\ref{ex:cpmm-vol}: The implied volatility as a function of the fixed leg $\bar\pi \in [0,2)$ with initial price $P_0^x = 1$ and swap maturity $T = 1$.}
\label{fig:cpmm-vol}
\end{subfigure}
~
\begin{subfigure}[t]{0.4\textwidth}
\centering
\begin{tikzpicture}[declare function={sx=2;sy=1;rho(\x)=(sx^2+sy^2-8*ln(2/(2-\x)))/(2*sx*sy);},]
\begin{axis}[
width={\textwidth},
xlabel={$\bar\pi$},
ylabel={$\rho^{CPMM}$},
xmin=0, xmax=2,
ymin=-1, ymax=1,
samples=200,
xticklabel={\pgfkeys{/pgf/number format/.cd,fixed,precision=2,zerofill}\pgfmathprintnumber{\tick}},
yticklabel={\pgfmathparse{\tick*100}\pgfmathprintnumber{\pgfmathresult}\%},
label style={font=\scriptsize},
ticklabel style={font=\tiny},
]
\addplot [domain={2*(1-exp(-1/8*(sx-sy)^2))}:{2*(1-exp(-1/8*(sx+sy)^2))},solid,color=black,very thick]{rho(x)};
\draw [dashed,color=black] ({2*(1-exp(-1/8*(sx-sy)^2))},-1) -- ({2*(1-exp(-1/8*(sx-sy)^2))},1);
\draw [dashed,color=black] ({2*(1-exp(-1/8*(sx+sy)^2))},-1) -- ({2*(1-exp(-1/8*(sx+sy)^2))},1);
\end{axis}
\end{tikzpicture}
\caption{Example~\ref{ex:cpmm-corr}: The implied correlation as a function of the fixed leg $\bar\pi \in [0.235,1.351]$ with initial prices $P_0^x = P_0^y = 1$, volatilities $\sigma_x = 200\%,~\sigma_y = 100\%$, and swap maturity $T = 1$.}
\label{fig:cpmm-corr}
\end{subfigure}
\caption{Visualizations of the implied volatility and implied correlation for the CPMM.}
\label{fig:cpmm}
\end{figure}

Following Theorem~\ref{thm:vol}, using pools in which one of the assets is the tokenized money market account, we are able to find all implied volatilities. 
Using these implied volatilities, we can use pools between two risky assets in order to compute the implied correlation $\rho^*$.
\begin{corollary}\label{cor:corr}
Consider an AMM between two assets with \emph{known} volatilities $\sigma_x,\sigma_y > 0$. Let $p_Z(\sigma) := \e{-\frac{\sigma^2}{2}T + \sigma\sqrt{T}Z} $ for $Z \sim N(0,1)$. Then for any $\bar\pi \in [C(P_0^x,P_0^y) - P_0^y \E[\bar C(P_0^x/P_0^y p_Z(|\sigma_x-\sigma_y|))] \, , \, C(P_0^x,P_0^y) - P_0^y \E[\bar C(P_0^x/P_0^y p_Z(\sigma_x+\sigma_y))]] \allowbreak \subseteq [0,C(P_0^x,P_0^y))$,
there exists a unique \emph{implied} correlation $\rho^* \in [-1,1]$ satisfying $\bar\pi = \E[\int_{0}^{T} e^{-rt} P_t^y \ell(P_t^x/P_t^y)dt]$.
\end{corollary}
\begin{remark}\label{rem:arb-corr}
In contrast to Remark~\ref{rem:arb-vol} in which we only need to consider a single pool of two assets, the arbitrage-free set of fixed leg payments for all combinations of \emph{three} assets is more complex. Within Corollary~\ref{cor:corr}, this is encoded within the provided price bounds 
$[C(P_0^x,P_0^y) - P_0^y \E[\bar C(P_0^x/P_0^y p_Z(|\sigma_x-\sigma_y|))] \, , \, C(P_0^x,P_0^y) - P_0^y \E[\bar C(P_0^x/P_0^y p_Z(\sigma_x+\sigma_y))]] \subseteq [0,C(P_0^x,P_0^y))$ which depends explicitly on the implied volatilities of two assets w.r.t.\ the same reference instrument (e.g., the money market account).
\end{remark}
\begin{example}\label{ex:cpmm-corr}
We conclude our CPMM case study of Example~\ref{ex:cpmm-vol} to now consider the implied correlation between two risky assets with known volatilities $\sigma_x,\sigma_y > 0$.
The implied correlation $\rho^*$ between these assets in the CPMM, given a price of $\bar\pi > 0$ for the fixed leg of the swap, is given by:
\[\rho^{CPMM} = \frac{\sigma_x^2 + \sigma_y^2 - \frac{8}{T}\log\left(\frac{2\sqrt{P_0^x P_0^y}}{2\sqrt{P_0^x P_0^y} - \bar\pi}\right)}{2\sigma_x\sigma_y}\]
for $\bar\pi \in 2\sqrt{P_0^x P_0^y} \times [1 - \exp(-\frac{1}{8}(\sigma_x - \sigma_y)^2 T) \; , \; 1 - \exp(-\frac{1}{8}(\sigma_x + \sigma_y)^2 T)] \subseteq [0,2\sqrt{P_0^x P_0^y})$. 
\end{example}

\begin{remark}
If this fixed-for-floating fee swap is introduced for multiple AMM structures simultaneously, it is possible that the implied volatilities and correlations are not identical.
This is because different AMMs concentrate their liquidity at different price ranges. In this way, a pseudo-implied volatility curve can be computed by comparing the volatilities quoted from different AMMs.
\end{remark}

\begin{remark}\label{rem:why}
While we have introduced this fixed-for-floating fee swap with the goal of providing implied volatilities and correlations, the participants have different incentives for partaking in this swap. 
Specifically, AMM liquidity providers auction off their share of the uncertain fee structure for an epoch; in exchange they are provided a certain return which can be a way to hedge the risks of this position.
In contrast, the investors that provide the fixed leg of the swap are able to earn the fees of the AMM without being subject to the risks associated with the rebalanced portfolio. For instance, the divergence or impermanent loss (see, e.g., \cite{xu2021sok,bichuch2022axioms}) is measured as the difference between the value of the initially provided assets of the AMM and the value of the updated portfolio within the AMM; by purchasing the uncertain fee stream only, the fixed leg investor is not exposed to this loss. Similar arguments can be made for loss-vs-rebalancing (\cite{milionis2022automated}).
\end{remark}

\section{Numerical Case Studies}\label{sec:cs}
In Section~\ref{sec:motivation}, we empirically found that the LVR has a strong linear relationship with the realized fees earned by LPs in the constant product market maker. Within this section, we wish to apply the results of the prior sections to investigate the reliability of fee-based volatility measures. In particular, we begin with an investigation of the simulated CPMM on SPY in Section~\ref{sec:cs-spy} in order to compare the fee volatility to reliable metrics for historical and implied volatilities. Noting that the fee volatility closely aligns with the historical volatility in this mature market as both are, by construct, backwards-looking, we are able to focus on that particular comparison in the remaining studies. Then, in Section~\ref{sec:cs-v3}, we return to the realized Uniswap v3 pool studied in Section~\ref{sec:motivation} to compare the fee and historical volatilities for a CPMM position. Finally, in Section~\ref{sec:cs-gld}, we compare two different simulated AMM designs (CPMM and Curve v1) on the GLD ETF to verify that the empirical results herein are AMM agnostic as theorized in Corollary~\ref{cor:cpmm}. For simplicity, throughout this section we take $r = 0$.

\subsection{S\&P 500: Historical and Implied Volatility}\label{sec:cs-spy}
Consider the simulated CPMM with 30bps fees on SPY data from 2023 introduced in Section~\ref{sec:motivation}. Recall that we collected bid-ask data at 1-second intervals and simulated transactions in the pool based solely on stale price arbitrage against the high bid and low ask during each interval. 
That is, at each second we allow an arbitrageur to use the highest bid (or lowest ask) to trade against the CPMM pool; notably, the arbitrageur only executes the trade if it is profitable when accounting for the 30bps fees assessed. With the simulated transactions, we compute the realized fees that would be earned by an LP. 
In this way, we are able to simulate the fees collected by an LP during this period. 

\begin{figure}[h!]
\centering
\begin{subfigure}[t]{0.45\textwidth}
\centering
\includegraphics[width=\textwidth]{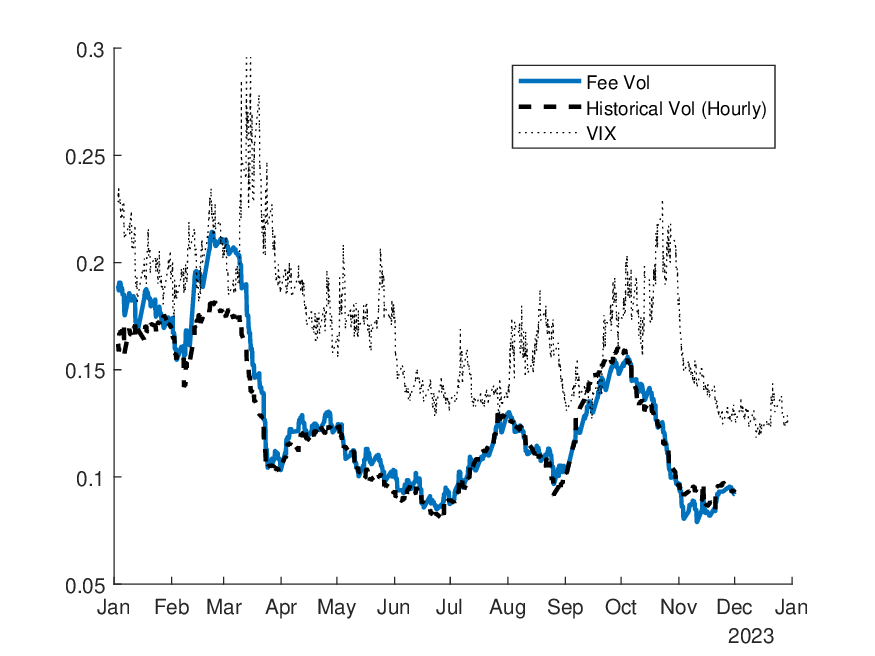}
\caption{Realized volatility from fees, together with historical and implied volatilities.}
\label{fig:spy-1}
\end{subfigure}
~
\begin{subfigure}[t]{0.45\textwidth}
\centering
\includegraphics[width=\textwidth]{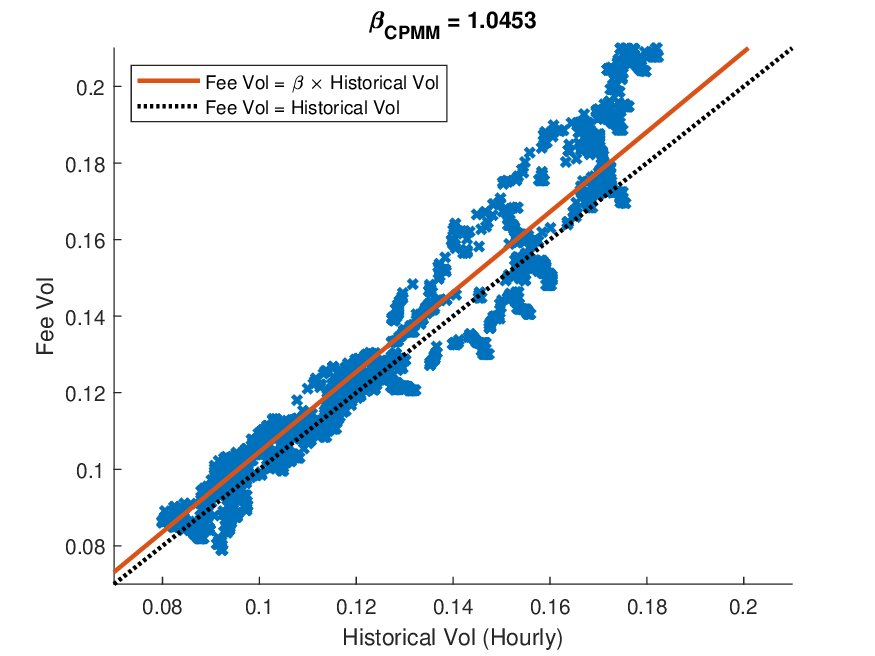}
\caption{Direct comparison of the fee and historical volatilities.}
\label{fig:spy-2}
\end{subfigure}
\caption{Realized fee, historical, and implied volatilities using SPY data.}
\end{figure}

As in Section~\ref{sec:motivation}, we aggregate the fees into overlapping 30-day intervals to evaluate the performance for an LP who has that specific time horizon. The 30-day horizon was chosen to align with the VIX which we use as a proxy for the 30-day \emph{implied volatility}. The \emph{historical volatility} over each 30-day window was computed using SPY hourly data (with the mid-price of the bid-ask spread) in the relevant period. Finally the \emph{fee volatility} was calculated along the lines of Theorem~\ref{thm:vol} and Example~\ref{ex:cpmm-vol}.
These three volatility measures are directly compared in Figure~\ref{fig:spy-1}. Immediately, we observe that the volatility calculated based on the realized fees is highly correlated to the realized historical volatility. Not surprisingly, both of these metrics are dominated by the forward-looking implied volatility (see., e.g., \cite{christensen1998relation,christensen2002new}). Notably, we expect that the fee volatility derived from an actual fixed-for-floating fee swap market, if such data existed, would more closely match the implied volatility than the historical volatility.
We further wish to note that, as observed in 2023, and as has been observed in traditional markets (see, e.g., \cite{christensen1998relation,christensen2002new}), the implied volatility typically overestimates the realized volatility; if the fixed-for-floating fee swap were to exist, this would lead to more stable and higher returns for the original LPs selling their fee stream.
Figure~\ref{fig:spy-2} highlights the strong linear relationship between the historical volatility and the fee volatility. As displayed in the title of that plot, we found that the fee volatility $\approx 1.0453 \times$ historical volatility.

\subsection{WETH/USDC Uniswap v3 Pool}\label{sec:cs-v3}

\begin{figure}[h!]
\centering
\begin{subfigure}[t]{0.45\textwidth}
\centering
\includegraphics[width=\textwidth]{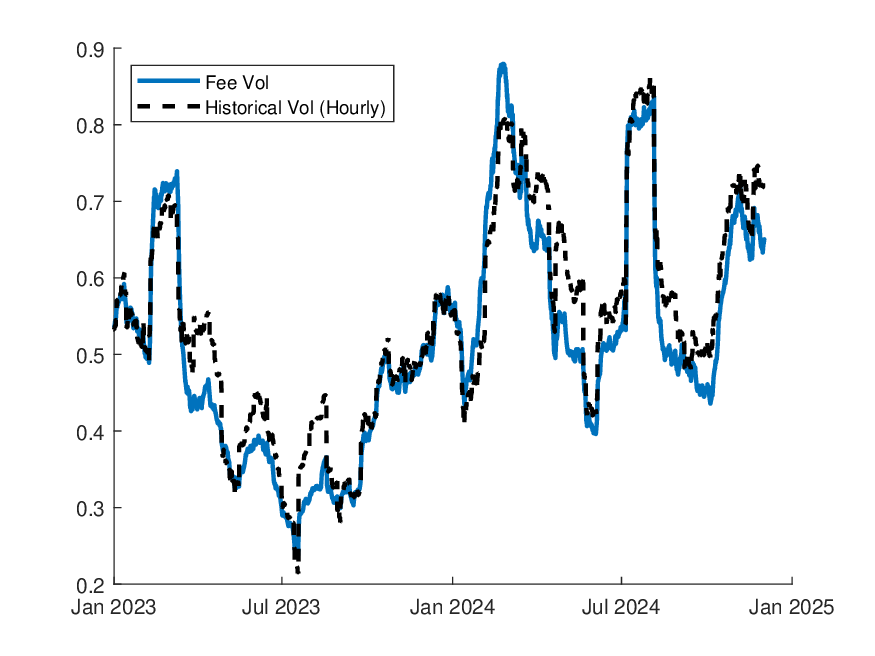}
\caption{Realized volatility from fees, together with the historical volatility.}
\label{fig:v3-1}
\end{subfigure}
~
\begin{subfigure}[t]{0.45\textwidth}
\centering
\includegraphics[width=\textwidth]{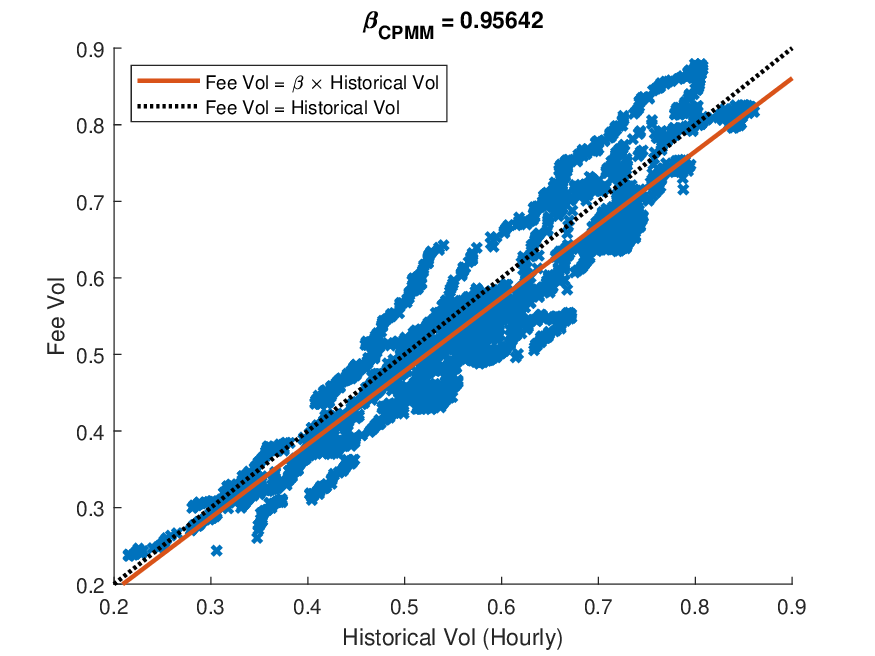}
\caption{Direct comparison of the fee and historical volatilities.}
\label{fig:v3-2}
\end{subfigure}
\caption{Realized fee and historical volatilities for a WETH/USDC Uniswap v3 pool.}
\label{fig:v3}
\end{figure}

While our prior example considered a simulated CPMM using data from traditional finance, we now proceed with actual AMM data from the same WETH/USDC (5bps fee) Uniswap v3 pool as used in Section~\ref{sec:motivation}. As in our prior use of this data, we consider an LP who invests over the entire price range to replicate a CPMM position. As in Section~\ref{sec:cs-spy}, we aggregate fees over a rolling window of 30 days and calculate the fee volatility following the methodology of Theorem~\ref{thm:vol} and Example~\ref{ex:cpmm-vol}. Additionally, we consider the 30-day historical volatility (from hourly spot prices) as a benchmark due to our findings on the simulated SPY data.
In Figure~\ref{fig:v3} we, again, find a strong relation between the fee volatility and the historical volatility. In fact, though the linear relation displayed between these volatilities has a slope of $0.95642$ (see Figure~\ref{fig:v3-2}), we can even recover a slope of $0.99791$ if we permit an intercept in the linear relationship. This confirms that the strong relationship observed in the simulated CPMM in Section~\ref{sec:cs-spy} holds true on actual DeFi data as well.\footnote{{In a limited study, we found that the implied volatility for WETH from crypto options data similarly overestimated both the fee and historical volatilities as observed in Section~\ref{sec:cs-spy} for SPY data.}}

\subsection{GLD: CPMM and Curve}\label{sec:cs-gld}

\begin{figure}[h!]
\centering
\begin{subfigure}[t]{0.45\textwidth}
\centering
\includegraphics[width=\textwidth]{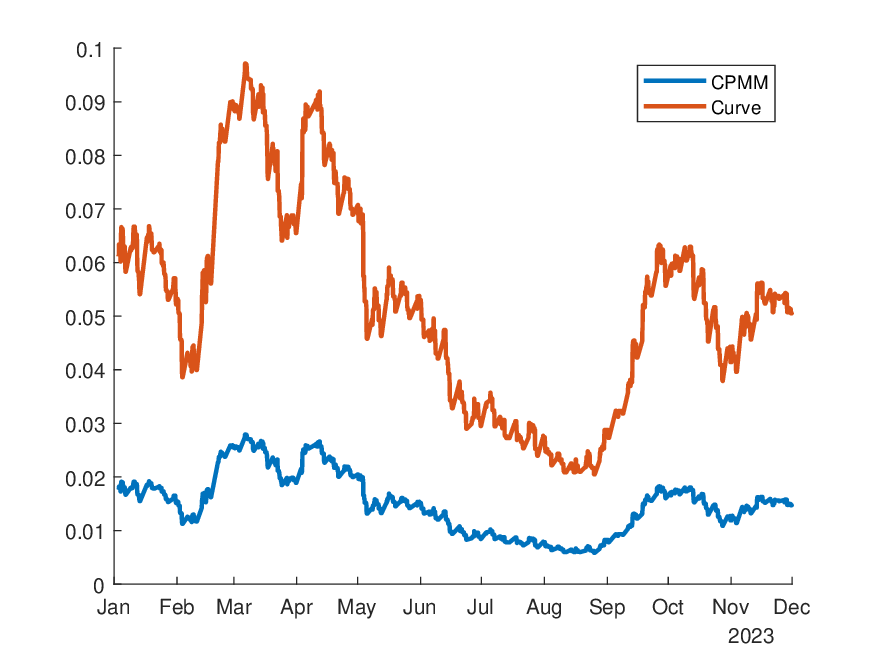}
\caption{Realized fees from a \$100 investment in CPMM and Curve v1 over 30-day rolling window.}
\label{fig:gld-fees}
\end{subfigure}
~
\begin{subfigure}[t]{0.45\textwidth}
\centering
\includegraphics[width=\textwidth]{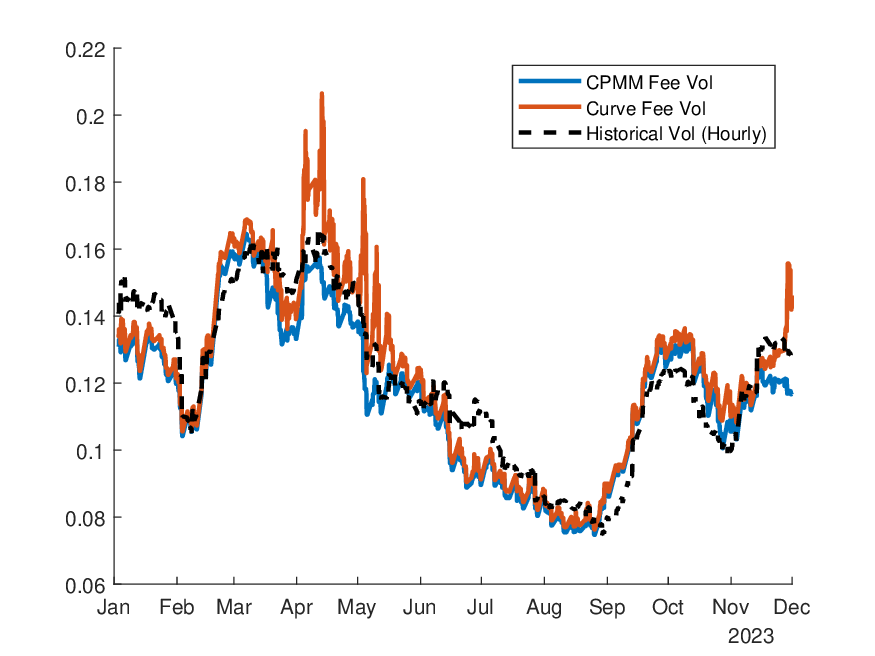}
\caption{Realized volatility from fees, together with the historical volatility.}
\label{fig:gld-vol}
\end{subfigure}
~
\begin{subfigure}[t]{0.45\textwidth}
\centering
\includegraphics[width=\textwidth]{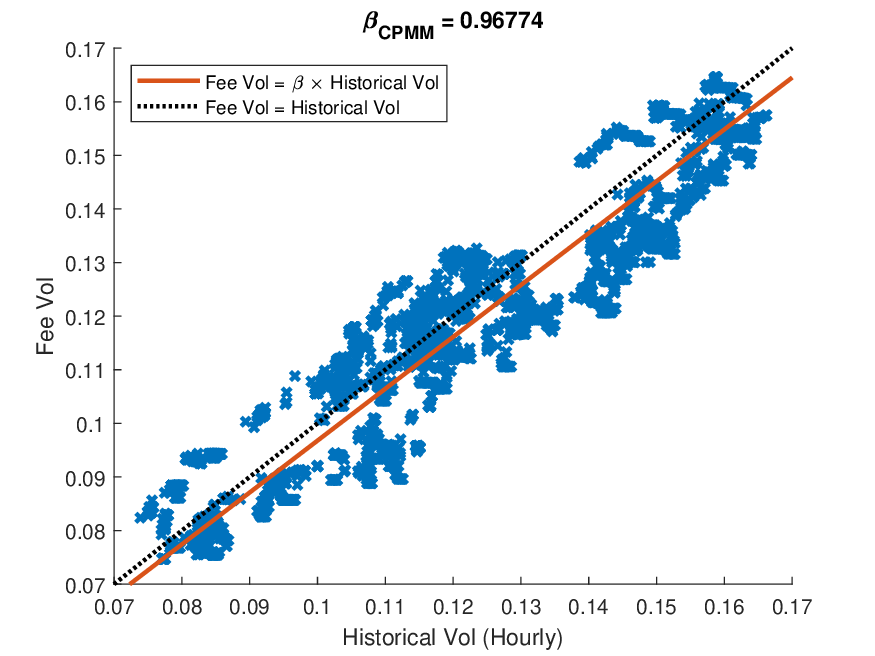}
\caption{Direct comparison of the fee and historical volatilities for CPMM.}
\label{fig:gld-cpmm}
\end{subfigure}
~
\begin{subfigure}[t]{0.45\textwidth}
\centering
\includegraphics[width=\textwidth]{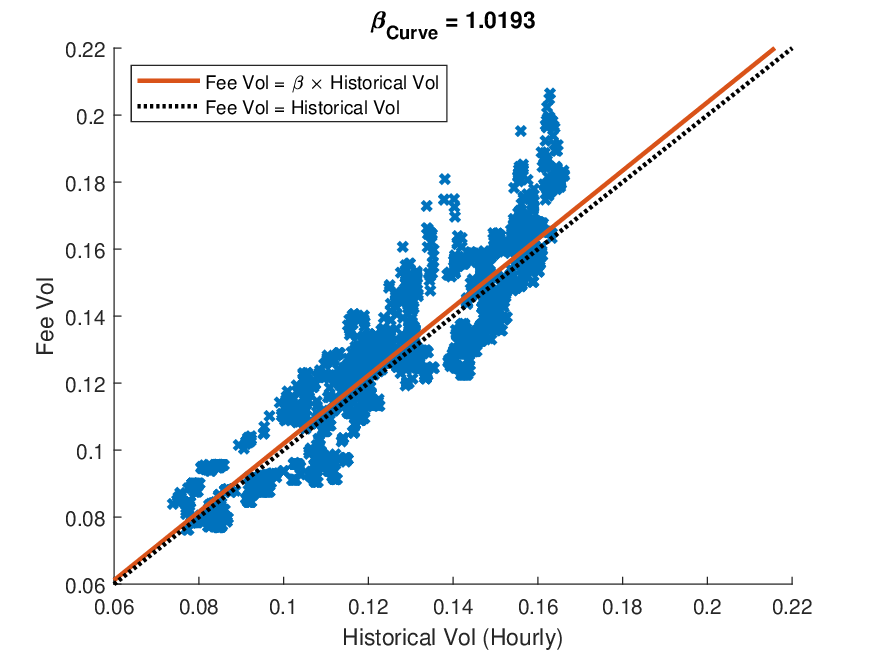}
\caption{Direct comparison of the fee and historical volatilities for Curve v1.}
\label{fig:gld-curve}
\end{subfigure}
\caption{Realized fee and historical volatilities using GLD data for CPMM and Curve v1 AMMs.}
\label{fig:gld}
\end{figure}

Our prior two case studies have focused entirely on the CPMM with both traditional and DeFi data. In this last case study, we want to consider a different AMM, namely Curve v1, which was designed to concentrate liquidity at a fixed price \cite{curve}. Because Curve was designed for assets that fluctuate around a consistent price, we chose to consider simulated AMMs on the GLD ETF, a commodity asset which should exhibit more mean-reverting properties, in 2023. The stable price for the Curve implementation was \$180 as it was close to the long-term average price during the period of study. With bid-ask data at 1-second intervals, we simulate both the CPMM and Curve pools based solely on stale price arbitrage against the high bid and low ask in each interval (as was done in Section~\ref{sec:cs-spy}). For both AMMs we assume a 30bps fee which accumulates from these arbitrage transactions; all fees are aggregated over a 30-day rolling window. As displayed in Figure~\ref{fig:gld-fees}, we see that the concentration of liquidity from Curve v1 leads to higher fee collections compared to the CPMM with the same initial investment. However, despite the different fee streams, we find that the fee volatilities from both the CPMM and Curve closely match the 30-day historical volatility (Figure~\ref{fig:gld-vol}).\footnote{We note that the Curve v1 fee volatility does not have a closed form, instead we use Monte Carlo simulations to calculate the expectations and a bisection search to find the unique volatility as in Theorem~\ref{thm:vol}.} The specific linear relationships between the historical and fee volatilities are displayed in Figures~\ref{fig:gld-cpmm} and~\ref{fig:gld-curve} for the CPMM and Curve respectively. These results empirically validate the theoretical equivalence of different AMM structures as presented in Corollary~\ref{cor:cpmm}.

\section{Conclusion}\label{sec:conclusion}

Within this work we analyzed the fee structures being paid by AMMs within a continuous-time framework. In particular, we found that the LVR is the unique risk-neutral fee structure which is consistent between all AMMs. 
With this fee structure, we propose a novel DeFi product -- a fixed-for-floating swap on the fees collected by the AMM. This swap can be used to attract more liquidity to decentralized exchanges as it allows for LPs to hedge the risky stream of fees for their liquidity position.
In fact, with this new instrument, implied volatilities and correlations can be quoted which would allow for more sophisticated strategies for investors. 
Though we believe this fixed-for-floating fee swap to be an important piece to the DeFi ecosystem, we leave it as future work to implement this new swap as a smart contract on blockchains. 
Empirically, we found that the realized fee-based volatility closely matches the historical volatility. We speculate that real pricing of the fixed-for-floating fee swap would more closely resemble the forward-looking implied volatility instead. However, only once these products are offered on-chain can such empirical studies be undertaken.

\bibliographystyle{plain}
\small{\bibliography{bibtex2}}

\appendix
\section{Proofs}\label{proofs}
\subsection{Proof of Lemma~\ref{lemma:amm}}
\begin{proof}
First, consider the AMM associated with the reachable set $R$. Define $(x(q),y(q))$ as selectors of $\argmin_{(x,y) \in R} (q x + y)$ for any $q > 0$. Trivially, $q x(q) + y(q) \leq q x(p) + y(p)$ for every $q,p > 0$. To prove monotonicity of $q \mapsto (x(q),y(q))$, fix $q' > q > 0$ and we consider three cases:
\begin{itemize}
\item Assume $x(q') > x(q)$ and $y(q') \geq y(q)$. Immediately, $q'x(q') + y(q') > q'x(q) + y(q)$ creating a contradiction to $(x(q'),y(q'))$ being minimizers.
\item Assume $x(q') \leq x(q)$ and $y(q') < y(q)$. Immediately, $qx(q) + y(q) > qx(q') + y(q')$ creating a contradiction to $(x(q),y(q))$ being minimizers.
\item Assume $x(q') > x(q)$ and $y(q') < y(q)$. Following from $(x(\cdot),y(\cdot))$ being minimizers, $q'x(q')+y(q')\leq q'x(q) + y(q)$ and $qx(q) + y(q)\leq qx(q')+y(q')$. By rearranging terms, we can conclude $y(q)-y(q') \leq q(x(q')-x(q)) < q'(x(q')-x(q)) \leq y(q)-y(q')$ providing a contradiction.
\end{itemize}
As none of these scenarios are possible, it must be that $x(q') \leq x(q)$ and $y(q') \geq y(q)$ as desired.

Now, consider $q \mapsto (x(q),y(q))$ and define the AMM by the reachable set $R = \bigcap_{q > 0} \{(\bar x,\bar y) \in \R^2_+ \; | \; q\bar x + \bar y \geq q x(q) + y(q)\}$. As the intersection of halfspaces and the nonnegative orthant, $R$ is trivially a closed, convex, upper set that is a subset of $\R^2_+$, i.e., $R = \cl\co(R + \R^2_+)$. Furthermore, since $q x(q) + y(q) \leq q x(p) + y(p)$ for every $q,p > 0$, it trivially follows that $(x(q),y(q)) \in R$ for every $q > 0$ proving that $R \neq \emptyset$.
\end{proof}

\subsection{Proof of Corollary~\ref{cor:amm-diff}}
\begin{proof}
This result trivially follows as the first property is equivalent to the desired monotonicity of the portfolio update functions and the second guarantees the minimality of $(x(q),y(q))$ at price $q$.
\end{proof}

\subsection{Proof of Proposition~\ref{prop:L}}
\begin{proof}
First, using $(x_{CPMM}^L(\cdot),y_{CPMM}^L(\cdot))$ provided in Example~\ref{ex:cpmm-xy}, we find that the curvature is given by $\kappa_{CPMM}^L(q) = -\frac{2q^{3/2}}{L (1 + q^2)^{3/2}}$.
Now, by the second property of the portfolio update functions in Corollary~\ref{cor:amm-diff}, we know that $y'(q) = -q x'(q)$ for every $q > 0$ and, therefore, $y''(q) = -x'(q) - q x''(q)$ as well.
Using this structure, we can simplify the curvature of any AMM as $\kappa(q) = -\left[(1 + q^2)^{3/2}x'(q)\right]^{-1}$.
Finally, we note that $\kappa(q) = \kappa_{CPMM}^L(q)$ if and only if $L = -2 q^{3/2}x'(q)$ as provided in~\eqref{eq:L} and the proof is completed.
\end{proof}

\subsection{Proof of Theorem~\ref{thm:fees}}
\begin{proof}
For simplicity, consider $V_t = \bar V_t + e^{-rt} C(P^x_t,P^y_t)$ where $\bar V_t := \int_0^t e^{-rt} F(P^x_t,P^y_t)dt$.
Immediately, by construction, $d\bar V_t = e^{-rt}F(P^x_t,P^y_t)dt$.
Furthermore, following an application of It\^o's lemma, and using the fact that $\bar C'(q) =x(q)$, we get: 
\begin{align*}
&d(e^{-rt}C(P^x_t,P^y_t)) \\
&\qquad= e^{-rt}\left[\begin{array}{l}\frac{1}{2}\left(\sigma_x^2 - \rho\sigma_x\sigma_y + \sigma_y^2\right)\left(\frac{(P^x_t)^2}{P^y_t}\right) x'(P^x_t/P^y_t)dt \\
         + \left(\sigma_x P^x_t x(P^x_t/P^y_t) + \rho\sigma_y P^y_t y(P^x_t/P^y_t)\right)dW_t + \sqrt{1-\rho^2}\sigma_y P^y_t y(P^x_t/P^y_t) d\bar{W}_t \end{array}\right].
\end{align*}
Recalling that $dV_t = d\bar V_t + d(e^{-rt}C(P^x_t,P^y_t))$, $V$ is a local martingale if and only if
\[e^{-rt}\left[F(p_x,p_y) + \frac{1}{2}\left(\sigma_x^2 - \rho\sigma_x\sigma_y + \sigma_y^2\right)\left(\frac{p_x^2}{p_y}\right) x'(p_x/p_y)\right] = 0\]
for every $(p_x,p_y) \in \R^2_{++}$. That is, $V$ is a local martingale if and only if $F(p_x,p_y) = p_y \ell(p_x/p_y)$ for every $(p_x,p_y) \in \R^2_{++}$.
The equivalent forms of $\ell$ follow from the construction of AMMs in Corollary~\ref{cor:amm-diff} so that $y'(q) = -q x'(q)$ for every $q > 0$.

To complete this proof, we now show that $V$ is, in fact, a true martingale.
By the Burkholder-Davis-Gundy inequality (see, e.g., \cite[Theorem 3.3.28]{karatzas2014brownian}), this result follows if $\E[\sqrt{\langle V \rangle_t}] \le \E[\langle V \rangle_t] < \infty$ for any $t\ge0$.
Due to the structure of the quadratic variation for $V$, this relies on having bounds on the growth conditions of $x(\cdot)$ and $y(\cdot)$. In fact, by concavity of $\bar{C}(\cdot) \geq 0$ (see Remark~\ref{rem:barC}), we will prove that $x(\cdot)$ and $y(\cdot)$ satisfy asymptotic growth conditions. Note that concavity of $\bar{C}$ trivially guarantees linear growth, i.e., $\bar{C}(q) \leq A + Bq$ for some $A,B > 0$ for every $q > 0$.
\begin{itemize}
\item \textbf{Linear Growth of $y$:} Following Assumption \ref{ass:differential} and Corollary \ref{cor:amm-diff}, $x(q)=\bar{C}'(q)$ is non-increasing and non-negative, so $x(q)$ is bounded at $\infty$. 
As a consequence, $y(q) = \bar{C}(q) - q x(q)$ also has at most linear growth at $\infty$. 
Since, by Corollary \ref{cor:amm-diff}, $y(\cdot)$ is non-decreasing and non-negative, it must satisfy a linear growth condition everywhere, i.e.,
$y(q) \le K_Y(1+q)$ for some $K_Y < \infty$ for every $q > 0$.
\item \textbf{Hyperbolic Growth of $x$:} By concavity, $\bar{C}'(q) \le \frac{\bar{C}(q) - \bar{C}(0)}{q-0}$ for any $q > 0$. Using the linear growth bound $\bar{C}(q) \le A + Bq$, we get:
$$
x(q) = \bar{C}'(q) \le \frac{A + Bq - \bar{C}(0)}{q} = \frac{\bar A}{q} + B,
$$
which shows that $x(q)$ can grow at most as $q^{-1}$ as $q \searrow 0$.
\end{itemize}

For notational convenience, let $q_t = P^x_t/P^y_t$. We can calculate the quadratic variation via
\begin{align*}
d\langle V \rangle_t = e^{-2rt} \left[ (x(q_t) P_t^x \sigma_x + y(q_t) P_t^y \sigma_y \rho)^2 + y^2(q_t) (P_t^y)^2 \sigma_y^2 (1-\rho^2) \right] dt.
\end{align*}
Using Fubini-Tonelli, we now check the finiteness of $\E[\langle V \rangle_t]$ for $t\ge0$ term by term. Beginning with the first and last term:
\begin{align*}
\E[\int_0^tx(q_s)^2 (P_s^x)^2ds] &\le \E\left[ \int_0^t\left(\frac{\bar A}{q_s} + B\right)^2 (P_s^x)^2ds \right] \le 2 \E\left[ \int_0^t\left(\frac{\bar A^2}{q_s^2} + B^2\right) (P_s^x)^2 ds\right]\\
&= 2\bar A^2 \int_0^t\E\left[ \frac{(P_s^x)^2}{q_s^2} \right]ds + 2B^2\int_0^t \E[(P_s^x)^2]ds\\
&= 2\bar A^2 \int_0^t\E\left[ (P_s^y)^2 \right]ds + 2B^2\int_0^t \E[(P_s^x)^2]ds<\infty,\\
\E[\int_0^t y(q_s)^2 (P_s^y)^2ds] &\le \E[K_Y^2 \int_0^t(1+q_s)^2 (P_s^y)^2ds] = K_Y^2 \E\left[ \int_0^t\left(1 + \frac{P_s^x}{P_s^y}\right)^2 (P_s^y)^2ds \right]\\
&=   K_Y^2 \int_0^t \E[(P_s^y)^2 + 2P_s^x P_s^y + (P_s^x)^2]ds<\infty,
\end{align*}
since all the terms are GBMs and their expectations are integrable.
Finally, consider the mixed term; applying the Cauchy-Schwarz inequality and recalling the finiteness of the first and last terms:
\begin{align*}
\E[\int_0^t x(q_s)y(q_s)P_s^x P_s^yds]  &= \int_0^t\E[ x(q_s)y(q_s)P_s^x P_s^y]ds \\
&\le \sqrt{\int_0^t\E[x(q_s)^2 (P_s^x)^2] ds }\sqrt{\int_0^t\E[y(q_s)^2 (P_s^y)^2] ds} < \infty.
\end{align*}
Therefore, $V_t$ is a local martingale with $\E[\langle V \rangle_t] < \infty$, which completes the proof.
\end{proof}

\subsection{Proof of Corollary~\ref{cor:indifference}}
\begin{proof}
Using the notation from Theorem~\ref{thm:fees}, $V(P^x_0,P^y_0) = \sup_{\tau\in\mathcal{T}}\E[V_\tau]$. Because $V$ is a martingale under the implied fee structure $(p_x,p_y) \in \R^2_{++} \mapsto p_y \ell(p_x/p_y)$, it follows that, for any choice of stopping time $\tau \in \mathcal{T}$, the same expectation is recovered and equals the cost of the liquidity position $C(P_0^x,P_0^y)$.
\end{proof}

\subsection{Proof of Corollary~\ref{cor:cpmm}}
\begin{proof}
First, $\ell_{CPMM}(q) = \frac{\sigma_x^2 - 2\rho\sigma_x\sigma_y + \sigma_y^2}{4}\sqrt{q}$ follows directly from $x_{CPMM}(q) = 1/\sqrt{q}$ and $y_{CPMM}(q) = \sqrt{q}$ for a single liquidity token of the CPMM.
With this structure, and noting that $L(q) = -2q^{3/2}x'(q)$ as defined in~\eqref{eq:L}, the equality $\ell(q) = L(q)\ell_{CPMM}(q)$ immediately follows.
\end{proof}

\subsection{Proof of Theorem~\ref{thm:vol}}
\begin{proof}
Recall that $C(p_x,p_y) = p_y \bar C(p_x/p_y)$ for any $p_x,p_y > 0$ with $\bar C(q) := q x(q) + y(q)$ for $q > 0$. Furthermore, we wish to recall from Remark~\ref{rem:barC} that $\bar C$ is nondecreasing and concave (in fact, $\bar C'(q) = x(q) \geq 0$ and $\bar C''(q) = x'(q) \leq 0$ for every $q > 0$ via Corollary~\ref{cor:amm-diff}).

By Theorem~\ref{thm:fees}, it follows that \[\bar C(P_0^x) = \E\left[\int_0^T \ell(e^{-rt}P_t^x) + \bar C(e^{-rT}P_T^x)\right].\]
Therefore, by rearranging terms, we find that \[\E\left[\int_0^T \ell(e^{-rt}P_t^x)\right] = \bar C(P_0^x) - \E[\bar C(e^{-rT}P_T^x)].\]
With this reformulation, we can prove that $\sigma\in\R_+\cup\{\infty\} \mapsto \E[\int_0^T \ell(e^{-rt} P_t^x)dt]$ is bijective with codomain $[0,\bar C(P_0^x)]$ by proving that $\sigma\in\R_+\mapsto \bar C(e^{-rT}P_T^x)$ is strictly decreasing with appropriate limits at $\sigma_x \in \{0,\infty\}$.
For simplicity of notation, define $p_Z(\sigma) := \exp(-\frac{\sigma^2}{2}T + \sigma\sqrt{T}Z)$ for random variable $Z$. Throughout this proof, let $Z,\bar{Z} \sim N(0,1)$ i.i.d.\ so that $e^{-rT} P_T^x(\sigma) \overset{(d)}{=} P_0^x p_Z(\sigma) \overset{(d)}{=} P_0^x p_{\bar Z}(\sigma)$ with dependencies on volatility explicit.

With this setting, we wish to prove three conditions: (i) $\E[\bar C(P_0^x p_Z(0))] = \bar C(P_0^x)$; (ii) $\sigma \mapsto \E[\bar C(P_0^x p_Z(\sigma))]$ strictly decreasing; and (iii) $\lim_{\sigma \to \infty} \E[\bar C(P_0^x p_Z(\sigma))] = 0$.
\begin{enumerate}
\item Note that $p_Z(0) = 1$ a.s.\ by construction. Immediately, the initial condition $\E[\bar C(P_0^x p_Z(0))] = \E[\bar C(P_0^x)] = \bar C(P_0^x)$ holds as desired.
\item Fix $\sigma_1,\sigma_2 > 0$. 
    \begin{align*}
    \E\left[\bar C(P_0^x p_Z(\sigma_1 + \sigma_2))\right] &= \E\left[\bar C\left(P_0^x p_Z(\sigma_1) p_{\bar Z}\left(\sqrt{\sigma_2^2 + 2\sigma_1\sigma_2}\right)\right)\right] \\
    &= \E\left[\E\left[\bar C\left(P_0^x p_Z(\sigma_1) p_{\bar Z}\left(\sqrt{\sigma_2^2 + 2\sigma_1\sigma_2}\right)\right) \; \Big | \; \bar Z\right]\right]\\ 
    &< \E\left[\bar C\left(\E\left[P_0^x p_Z(\sigma_1) p_{\bar Z}\left(\sqrt{\sigma_2^2 + 2\sigma_1\sigma_2}\right) \; \Big | \; \bar Z\right]\right)\right]\\
    &= \E\left[\bar C(P_0^x p_Z(\sigma_1))\right]
    \end{align*}
    The inequality follows from Jensen's inequality and Assumption~\ref{ass:strict} so that it is a strict inequality. Therefore $\sigma \mapsto \E[\bar C(e^{-rt}P_t^x(\sigma))]$ is strictly decreasing. 
\item By monotone convergence and the above results, $\lim_{\sigma \to \infty} \E[\bar C(P_0^x p_Z(\sigma))] \in [0,\bar C(P_0^x))$ exists since $\bar C(q) \geq 0$ by construction for any $q > 0$.  We wish to prove that this limit is 0.
We note that $\lim_{q \to \infty} \bar C(q)/q = 0$ by Assumption~\ref{ass:R-bd} (either $\bar C(q) \leq \bar c \in \R$ for every $q > 0$ or if $\lim_{q \to \infty} \bar C(q)=\infty$ then  $\lim_{q \to \infty} \bar C(q)/q = \lim_{q \to \infty} x(q) = 0$ by L'H\^opital's rule). 
Consider $\hat C(p) := x(1/p) + p y(1/p)$ for $p > 0$ to be the pool value function under the change of num\'eraire (i.e., with $x$ as the num\'eraire).
Therefore, by symmetry between the assets,
\begin{align*}
0 &= \lim_{p \to \infty} \hat C(p)/p = \lim_{p \to \infty} [x(1/p)/p + y(1/p)] = \lim_{q \searrow 0} [q x(q) + y(q)] = \lim_{q \searrow 0} \bar C(q).
\end{align*}

With this result, we directly want to consider $\E[\bar C(P_0^x p_Z(\sigma))]$ in order to determine its limit.
Fix $\epsilon > 0$ arbitrary. Let $0 < \delta \leq \Delta$ such that $\bar C(q) \leq \epsilon$ for every $q < \delta$ and $\bar C(q) \leq \epsilon q$ for every $q > \Delta$.\footnote{Existence of $\delta,\Delta$ follow from $\lim_{q \searrow 0} \bar C(q) = 0$ and $\lim_{q \to \infty} \bar C(q)/q = 0$ respectively.}
We now partition the price space based on $\delta,\Delta$, i.e.,
\begin{align*}
\E[\bar C(P_0^x p_Z(\sigma)] &= \underbrace{\E[\bar C(P_0^x p_Z(\sigma)) \ind{P_0^x p_Z(\sigma) < \delta}]}_{=: I_1(\sigma)} + \underbrace{\E[\bar C(P_0^x p_Z(\sigma)) \ind{P_0^x p_Z(\sigma) \in [\delta,\Delta]}]}_{=: I_2(\sigma)} \\
&+ \underbrace{\E[\bar C(P_0^x p_Z(\sigma)) \ind{P_0^x p_Z(\sigma) > \Delta}]}_{=: I_3(\sigma)}.
\end{align*}
With this partitioning, we wish to consider the limiting behavior $\limsup\limits_{\sigma \to \infty} I_1(\sigma)$, $\limsup\limits_{\sigma \to \infty} I_2(\sigma)$, and $\limsup\limits_{\sigma \to \infty} I_3(\sigma)$:
\begin{align*}
\limsup_{\sigma\to\infty} I_1(\sigma) &\leq \epsilon \limsup_{\sigma\to\infty} \P(P_0^x p_Z(\sigma) < \delta) 
    \le \epsilon, \\
\limsup_{\sigma\to\infty} I_2(\sigma) &\leq \bar C(\Delta) \limsup_{\sigma\to\infty} \P(P_0^x p_Z(\sigma) \in [\delta,\Delta])\\ 
    &= \bar C(\Delta) \limsup_{\sigma\to\infty} \left[\Phi\left(\frac{\log(\Delta/P_0^x)+\frac{\sigma^2}{2}T}{\sigma\sqrt{T}}\right) - \Phi\left(\frac{\log(\delta/P_0^x)+\frac{\sigma^2}{2}T}{\sigma\sqrt{T}}\right)\right] = 0, \\
\limsup_{\sigma\to\infty} I_3(\sigma) &= \limsup_{\sigma\to\infty}\E\left[P_0^x p_Z(\sigma) \left(\frac{\bar C(P_0^x p_Z(\sigma))}{P_0^x p_Z(\sigma)}\right) \ind{P_0^x p_Z(\sigma) > \Delta}\right]\\
    &\leq \limsup_{\sigma\to\infty}P_0^x \epsilon \E[p_Z(\sigma) \ind{P_0^x p_Z(\sigma) > \Delta}]\\
    &= P_0^x \epsilon \limsup_{\sigma \to \infty} \Phi\left(-\frac{\log(\Delta/P_0^x) - \frac{\sigma^2}{2}T}{\sigma\sqrt{T}}\right) = P_0^x \epsilon.
\end{align*}
That is, $\E[\bar C(P_0^x p_Z(\sigma)] \leq \limsup\limits_{\sigma\to\infty} I_1(\sigma) + \limsup\limits_{\sigma\to\infty} I_2(\sigma) + \limsup\limits_{\sigma\to\infty} I_3(\sigma) \leq (1 + P_0^x)\epsilon$ for any $\epsilon > 0$ and the desired result is proven.
\end{enumerate}
\end{proof}

\subsection{Proof of Corollary~\ref{cor:corr}}
\begin{proof}
Following the same logic as in the proof of Theorem~\ref{thm:vol}, we note that $\E[\int_0^T e^{-rt} P_t^y \ell(P_t^x/P_t^y)dt] = C(P_0^x,P_0^y) - \E[e^{-rT}P_T^y \bar C(P_T^x/P_T^y)]$.
However, instead of considering the expectation $\E[e^{-rT} P_T^y \bar C(P_T^x/P_T^y)]$ under the $\P$-measure, we conduct the change of measure $\frac{d\Q}{d\P} = e^{-rT}P_T^y/P_0^y$.
Therefore, we find \[\E[e^{-rT}P_T^y \bar C(P_T^x/P_T^y)] = P_0^y \E^\Q[\bar C(P_T^x/P_T^y)] = P_0^y\E^\Q[\bar C(P_0^x/P_0^y p_Z(\sqrt{\sigma_x^2 - 2\rho\sigma_x\sigma_y + \sigma_y^2})) ]\]
where $Z \sim N(0,1)$ under $\Q$.
Therefore, as follows from Theorem~\ref{thm:vol}, we have the bijection between $\bar\sigma := \sqrt{\sigma_x^x - 2\rho\sigma_x\sigma_y + \sigma_y^2} \geq 0$ and $\bar\pi \in [0,C(P_0^x,P_0^y))$.
The proof can be concluded by solving $\rho^* = \frac{\sigma_x^2 + \sigma_y^2 - \bar\sigma^2}{2\sigma_x\sigma_y}$ and restricting the relevant fixed costs $\bar\pi$ so that $\rho^* \in [-1,1]$ (i.e., so that $\bar\sigma \in [|\sigma_x-\sigma_y| , \sigma_x+\sigma_y]$).
\end{proof}

\end{document}